\documentclass{article}

\usepackage[final]{neurips_2024}
\usepackage{graphicx}

\usepackage[utf8]{inputenc} 
\usepackage[T1]{fontenc}    
\usepackage{hyperref}       
\usepackage{url}            
\usepackage{booktabs}       
\usepackage{amsfonts}       
\usepackage{nicefrac}       
\usepackage{microtype}      
\usepackage{xcolor}         
\usepackage{amsmath}

\title{Musical Agent Systems: MACAT and MACataRT}

\newcommand*\samethanks[1][\value{footnote}]{\footnotemark[#1]}
\author{
  Keon Ju M. Lee\thanks{Metacreation Lab for Creative AI: \url{https://www.metacreation.net/}} \\
  School of Interactive Arts and Technology\\
  Simon Fraser University\\
  Vancouver, British Columbia, Canada \\
  \texttt{keon\_maverick@sfu.ca} \\
  \And
  Philippe Pasquier\samethanks \\
  School of Interactive Arts and Technology \\
  Simon Fraser University \\
  Vancouver, British Columbia, Canada \\
  \texttt{pasquier@sfu.ca} \\
}

\begin{document}

\maketitle

\begin{abstract}
Our research explores the development and application of musical agents, human-in-the-loop generative AI systems designed to support music performance and improvisation within co-creative spaces. We introduce MACAT and MACataRT, two distinct musical agent systems crafted to enhance interactive music-making between human musicians and AI. MACAT is optimized for agent-led performance, employing real-time synthesis and self-listening to shape its output autonomously, while MACataRT provides a flexible environment for collaborative improvisation through audio mosaicing and sequence-based learning. Both systems emphasize training on personalized, small datasets, fostering ethical and transparent AI engagement that respects artistic integrity. This research highlights how interactive, artist-centred generative AI can expand creative possibilities, empowering musicians to explore new forms of artistic expression in real-time, performance-driven and music improvisation contexts.
\end{abstract}

\section{Introduction}
\subsection{Background}
A \textbf{musical agent} [1] is an artificial system designed to automate creative musical tasks within the field of \textbf{musical metacreation}—an area focused on the computational simulation of musical creativity [2]. In this domain, musical agents perform a variety of crucial tasks that deepen and broaden human-AI collaboration in music-making. These tasks encompass generative composition, where agents autonomously produce original material based on learned styles; interactive performance, which allows agents to adapt in real-time to live inputs from human musicians; and accompaniment, in which agents provide dynamic, context-sensitive support that complements primary musical performances. Further, style adaptation enables agents to align their outputs with specific genres or artist preferences. At the same time, real-time improvisation allows for spontaneous, unscripted musical responses, fostering a creative and responsive interaction with human musicians. These tasks position musical agents as versatile, adaptive collaborators, functioning as co-creators, responsive partners, and innovative contributors across diverse musical contexts. 

Drawing on principles from Artificial Intelligence (AI [3]) and Multi-Agent Systems (MAS [4]), musical agents operate autonomously, making decisions and performing actions in response to real-time musical contexts. These systems range from simple rule-based models to advanced frameworks capable of learning and adapting through interaction. Often implemented using MAX/MSP programming [5], they facilitate collaboration with human musicians or other agents, showcasing attributes such as reactivity, adaptability, and coordination. Complementing these agents, Corpus-Based Concatenative Synthesis (CBCS)—pioneered in IRCAM's CataRT system [6] and widely used in electroacoustic music [7,8,9]—utilizes small, personalized datasets of recordings to generate new content. CBCS achieves this by selecting and transforming audio segments from a pre-existing corpus, enabling real-time control of the generative process while preserving the expressive nuances of human performance.

\subsection{Research Motivation}
Our research is motivated by the aim of developing interactive, artist-in-the-loop generative AI systems designed for musicians and sound artists. The primary objective is to enable musicians to explore novel creative possibilities and expand their musical and artistic practices, particularly within real-time scenarios and interactive environments. 

Our musical agent systems, MACAT and MACataRT, are based on the mindset of using small data in music [10] and the principles of model crafting in visual arts with generative AI [11], as well as fostering music-making in co-creative spaces between human musicians and musical agents [12], 
including machine listening [13].

Training on a small, high-quality audio corpus enables our musical agent systems to closely align with the specific musical nuances and stylistic preferences of collaborating musicians. In contrast to models that broadly reproduce generalized patterns derived from large-scale audio or MIDI datasets, our model is optimized to interpret and respond to the unique attributes embedded within a musician's work, thereby facilitating the personalization of music systems. This focused approach enhances the agent's capacity to function as a genuine creative collaborator, fostering stylistic coherence and adaptability that could otherwise be diluted in models trained on more generic, large-scale audio or MIDI corpora.

Our musical agent systems are highly flexible, enabling artists to train the models using curated datasets tailored specifically to their unique stylistic preferences. This approach, which we term "model crafting," allows artists to incorporate their own recorded data into the training process, creating an agent that is not bound by predefined genres or styles but is instead highly personalized to their artistic identity. By selecting and shaping the training corpus, artists can ensure that the musical agent aligns closely with their aesthetic vision, resulting in outputs reflecting their individual voices and creative nuances. This flexibility makes the system adaptable to a wide range of musical expressions, empowering artists to use AI not as a generic accompaniment tool but as a customizable co-creator that can evolve with their artistic practice.

\section{Workflow of Musical Agent Systems}
\begin{figure}[!ht]
\centering
\includegraphics[width=0.85\columnwidth]{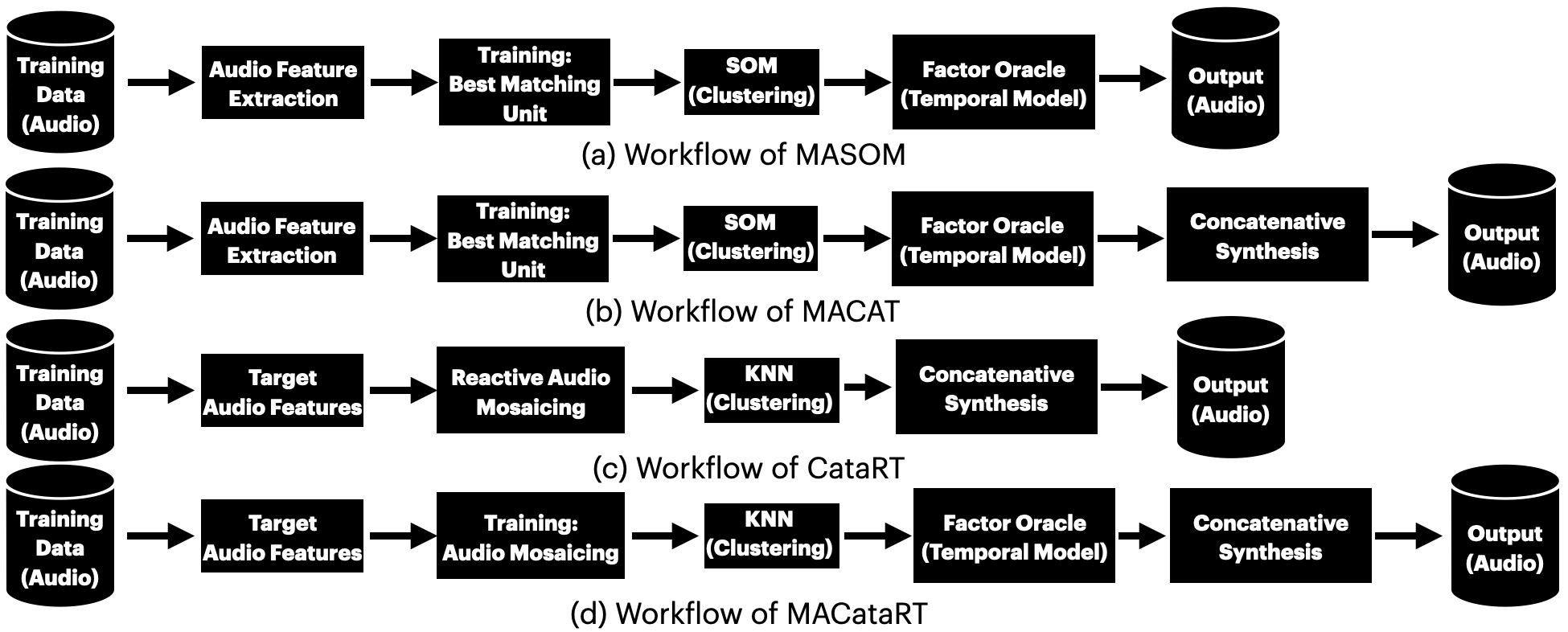}
 \caption{The workflow of each musical agent system for the comparison: (a) MASOM, (b) MACAT, (c) CataRT, and (d) MACataRT.}
 \label{fig:musical-agent-workflow}
\end{figure}
\subsection{Workflow of MASOM and MACAT}
Musical Agent based on Self-Organizing Maps (MASOM [14]) is a machine improvisation system for live performances, particularly suited for experimental music and free improvisation. It integrates self-organizing maps (SOM [15]) as a sound memory, Variable Markov Models (VMM [16]) to recognize and generate musical structures, and affective computing for real-time audio analysis. MASOM can listen to and generate audio signals, learning from a corpus of recordings to structure its performances. Appendix B provides a comprehensive explanation of the latest version of MASOM.

\textbf{MACAT}, as shown in the workflow in Figure \ref{fig:musical-agent-workflow} (b), is an enhanced version of the MASOM agent developed by the Metacreation Lab, featuring integrated real-time sound synthesis and visualization. At its core, MACAT utilizes concatenative sound synthesis and offers improved visualization for past and current nodes. It enables sound artists to flexibly synthesize audio segments and explore a broad range of timbral possibilities through sound design parameters. It employs the SOM and Factor Oracle (FO [17]), a suffix automaton for real-time pattern recognition in sequences of nodes representing clusters of audio segments grouped by timbral similarity. During training, MACAT conducts offline machine listening to analyze the original audio data, initialize SOM nodes, and identify the Best Matching Unit (BMU) for each input vector based on Euclidean distance. The BMU and its neighbours are adjusted, resulting in a SOM that clusters audio segments into groups. Subsequently, MACAT learns the sequence of nodes using a VMM, with the FO identifying patterns within this sequence during real-time generation. Dynamically adapting to its self-listening process, MACAT uses a congruence parameter to regulate FO’s forward and backward transitions, ensuring responsive and context-sensitive generative output for live performances and creative exploration.

\subsection{Workflow of CataRT and MACataRT}
The IRCAM's CataRT system [6] is a real-time corpus-based concatenative synthesis tool designed for interactive sound exploration by selecting sound units from a database based on audio descriptors. Implemented in Max/MSP, the system organizes sounds within an audio descriptor space, enabling users to target specific audio features, such as pitch or timbre, as illustrated in the workflow of Figure \ref{fig:musical-agent-workflow} (c). Building on traditional granular synthesis, the system utilizes a curated corpus of segmented audio crafted by artists, enabling precise control over sound characteristics through concatenative sound synthesis. Its versatile applications span interactive sound synthesis, gesture-controlled synthesis, live audio resynthesis, and expressive speech synthesis, offering a flexible and powerful interface for creative sound exploration.

\textbf{MACataRT}, as depicted in the workflow in Figure \ref{fig:musical-agent-workflow} (d), is an enhanced version of the CataRT system, incorporating a temporal model based on the factor oracle and offering a more intuitive interface for sound synthesis and resampling. The original CataRT system, while capable of reactive improvisation through real-time machine listening, clusters audio segments using K-Nearest Neighbors (KNN [18]) and targets specific audio features for audio mosaicing. However, it lacks a temporal model to manage time-based musical structures. To address this limitation, MACataRT integrates the factor oracle to automate the generation process, building on CataRT's audio mosaicing capability. Audio mosaicing [19] assembles audio segments into new audio pieces by selecting segments that match target audio features specified by musicians, who then refine the output to replicate the desired characteristics. MACataRT enhances this process with interactive audio mosaicing that functions in both real-time and offline modes. In real-time, the musical agent facilitates reactive improvisation, responding to live inputs based on machine listening and targeting audio features without using the factor oracle. In its proactive improvisation mode, the musical agent system learns sequences of audio segment indices during offline training, enabling the factor oracle to generate music based on these learned sequences. This dual capability allows musicians to interactively play alongside MACataRT, fostering dynamic, creative exchanges. Details on the interface of MACAT and MACataRT systems are provided in Appendix C.
\section{Research-Creation and Musical Practice}
\subsection{Research-Creation Methodology}

The research-creation methodology in musical practice [20], especially in performances with musical AI agents, combines scholarly inquiry with creative experimentation, focusing on co-creation between humans and AI systems. Musicians interact with AI agents as collaborative partners, capable of responding to real-time inputs and aligning with the performer’s expressive intent. This relationship fosters a dynamic musical dialogue where AI agents generate, adapt, and influence live performances, broadening the possibilities for improvisation, stylistic adaptation, and spontaneous composition. This approach merges artistic exploration with technical research, offering insights into AI’s creative potential in music and redefining the musician’s role in AI-augmented performances.

The research-creation methodology is essential in musical performance and improvisation, enabling a deeper and more nuanced exploration of the experiential and artistic aspects of music-making that quantitative analysis alone cannot fully capture. In such contexts, emphasis is placed on interpretive, spontaneous, and affective qualities that emerge through real-time collaboration, particularly with musical AI agents. While quantitative analysis offers insights into measurable elements like timing accuracy and frequency distribution, it fails to address the subjective, responsive, and context-sensitive dimensions of co-creation that are fundamental to live musical practice. Musical agent systems, regarded as intelligent musical instruments, require research-creation to cultivate these interactive relationships, allowing artists to develop virtuosity in collaboration with AI, create aesthetically compelling music, and explore new or unique musical styles. Given the absence of well-established quantitative measures for evaluating personalized improvisation systems and diverse musical scenarios, our approach prioritizes showcasing musical practice and performance over suboptimal computational and quantitative evaluation.

\subsection{Musical Application and Artistic Practice}
The application of each musical agent varies, offering musical affordances that support the research-creation in real-world musical performance scenarios. MACAT, for instance, is particularly effective in solo performance settings, where its improvisational output adapts through a self-listening process that allows it to take the lead in most musical contexts. MACAT was showcased at the MusicAcoustica Festival in Hangzhou, China, by the artist collective K-Phi-A, featuring Philippe Pasquier on live ambient electronics, Keon Ju Maverick Lee on percussions, and VJ Amagi providing audiovisuals.

MACataRT expands creative possibilities through audio mosaicing, supporting both reactive and proactive improvisation and proving its adaptability across diverse collaborative settings. Its practical effectiveness was demonstrated in a live performance, where it was used by a percussionist (Keon) and a guitarist (Sara) to co-create music. This collaboration earned significant recognition when their piece, \textit{Echoes of Synthetic Forest} by the music duo \textit{KeRa}, was selected as one of the Top 10 finalists in the 2024 AI Music Song Contest [21] and performed in Zürich, Switzerland [22]. These achievements highlight the value of research-creation, showcasing how musical agents can enhance real-time artistic expression and foster collaborative exploration. Our musical agent systems and showcases are available for public access on the Metacreation Lab's GitHub repository\footnote{\url{https://github.com/Metacreation-Lab/Musical-Agent-Systems}}.

\section{Conclusion and Future Work}

Our research highlights the potential of the musical agents MACAT and MACataRT to enable creative collaboration between AI and human musicians, offering a novel approach to interactive music-making through real-time generative AI. These systems prioritize the preservation of expressive nuances by employing corpus-based concatenative synthesis and small-data training methods, allowing the musical agents to act as responsive co-creators in diverse performance contexts. MACAT and MACataRT demonstrate how artist-in-the-loop AI agents can significantly broaden the creative options for musicians, providing practical tools that integrate into live performances and improvisational settings. Ethical considerations related to our musical agents are discussed in Appendix A.

In future work, we aim to enhance both the temporality and explainability of our musical agent systems. Currently, these systems utilize conventional audio mosaicing techniques and the factor oracle algorithm for temporal modelling to identify and generate musical patterns. To improve temporality, we plan to integrate deep learning architectures that enable agents to learn and retain longer sequences of musical patterns. For enhanced explainability, we intend to incorporate a module that records the history of past musical patterns, thereby advancing comprehension from a note-level to a bar-level understanding. Additionally, we aim to advance the machine listening module to deepen the agents' musical comprehension. A feedback loop using reinforcement learning may also be introduced to further enhance the adaptability of musical agents in real-time performance contexts.

\section*{References}

{
\small

[1] Tatar, K., \& Pasquier, P. (2019). Musical agents: A typology and state of the art towards musical metacreation. Journal of New Music Research.

[2] Pasquier, P., Eigenfeldt, A., Bown, O., \& Dubnov, S. (2017). An introduction to musical metacreation. Computers in Entertainment (CIE)

[3] Russell, S. J., \& Norvig, P. (2016). Artificial intelligence: a modern approach. Pearson.

[4] Wooldridge, M. (2009). An introduction to multiagent systems. John Wiley \& sons.

[5] Puckette, M. (1991). Combining event and signal processing in the MAX graphical programming environment. Computer music journal.

[6] Schwarz, D., Beller, G., Verbrugghe, B., \& Britton, S. (2006). Real-time corpus-based concatenative synthesis with CataRT. In the proceedings of the Conference on Digital Audio Effects (DAFx).

[7] Sturm, B. L. (2006). Adaptive Concatenative Sound Synthesis and Its Application to Micromontage Compositior. Computer Music Journal.

[8] Hackbarth, B., Schnell, N., Esling, P., \& Schwarz, D. (2013). Composing morphology: Concatenative synthesis as an intuitive medium for prescribing sound in time. Contemporary Music Review.

[9] Einbond, A., Bresson, J., Schwarz, D., \& Carpentier, T. (2021). Instrumental Radiation Patterns as Models for Corpus-Based Spatial Sound Synthesis: Cosmologies for Piano and 3D Electronics. In the proceedings of the International Computer Music Conference (ICMC).

[10] Vigliensoni, G., Perry, P., \& Fiebrink, R. (2022). A small-data mindset for generative AI creative work. In the proceedings of the Machine Learning for Creativity and Design Workshop at the conference on Neural Information Processing Systems (NIPS).

[11] Abuzuraiq, A. M., \& Pasquier, P. (2024). Seizing the Means of Production: Exploring the Landscape of Crafting, Adapting and Navigating Generative AI Models in the Visual Arts. In the proceedings of the Computer-Human Interaction (CHI) workshop on Generative AI and HCI.

[12] Thelle, N. J. W., \& Wærstad, B. I. G. (2023). Co-Creatives Spaces: The machine as a collaborator. In Proceedings of the Conference on New Interfaces for Musical Expression (NIME).

[13] Rowe, R. (1992). Interactive music systems: machine listening and composing. MIT press.

[14] Tatar, K. \& Pasquier, P. (2017). MASOM: A Musical Agent Architecture based on Self-Organizing Maps, Affective Computing, and Variable Markov Models. In the proceedings of the Conference on International Workshop on Musical Metacreation (MuMe).

[15] Kohonen, T. (1990). The self-organizing map. In the proceedings of the conference on the Institute of Electrical and Electronics Engineers (IEEE).

[16] Begleiter, R., El-Yaniv, R., \& Yona, G. (2004). On prediction using variable order Markov models. Journal of Artificial Intelligence Research

[17] Assayag, G., \& Dubnov, S. (2004). Using factor oracles for machine improvisation. Journal of Soft Computing.
 
[18] Cover, T., \& Hart, P. (1967). Nearest neighbor pattern classification. IEEE transactions on information theory.

[19] Lazier, A., \& Cook, P. (2003). MOSIEVIUS: Feature driven interactive audio mosaicing. In the proceedings of the conference on Digital audio effects (DAFx).

[20]Stévance, S., \& Lacasse, S. (2017). creation in Music and the Arts: Towards a Collaborative Interdiscipline. Routledge.

[21] Top 10 Finalist AI Music Song Contest (2024) \\ URL:  \url{https://www.aisongcontest.com/the-2024-finalists} (Last accessed: 14th of November, 2024).

[22] KeRa's Music Performance as a Top 10 Finalist at Zürich, Switzerland (2024) \\ URL: \url{https://www.youtube.com/live/2ojQEGNUXic?si=nEmzbR_yCyhyk1aS&t=4499} (Last accessed: 14th of November, 2024).


\clearpage
\appendix
\section{Ethical Implication}
Our musical agent systems present multiple advantages over large-scale generative AI models, particularly regarding ethical considerations and data transparency. In contrast to big data models, which often lack traceability and explainability regarding how their training data influence generated outputs—thereby risking the incorporation of stylistic elements from other musicians without consent—our systems utilize small, personalized datasets of music recordings created by specific musicians. This strategy not only promotes more ethical use of AI in generative music by respecting the intellectual property rights and artistic contributions of other composers but also enhances the transparency and accountability of the creative process.

By employing small, personalized datasets and well-documented training data, our systems facilitate straightforward tracking of how specific pieces of music contribute to and shape the generated content. This approach provides a clearer explanation of the AI's creative influences, ensuring that the resulting music is original and ethically produced. Additionally, using curated, smaller datasets reduces the likelihood of inadvertently mimicking the styles or techniques of musicians who have not explicitly agreed to contribute to the training process, thereby maintaining artistic integrity and reducing the potential for ethical breaches related to appropriation or unintentional plagiarism.

Moreover, the reliance on small, personalized datasets enables more precise control over the creative direction of the AI's outputs. It allows musicians and composers to tailor the generative process to align with their artistic vision and preferences, resulting in a more meaningful collaboration between human creativity and AI technology. This level of customization is often unattainable in large-scale generative models, where the vastness and anonymity of the data pool obscure the specific influences that guide the AI's creative decisions. The transparency afforded by our approach enhances the accountability of AI-generated music. It allows for a clear audit trail, where the contributions of individual pieces to the final output can be explicitly identified and verified. This feature is crucial in contexts where the ethical implications of AI usage are scrutinized, such as in academic research, commercial music production, or public performances. By maintaining a transparent and ethical framework, our musical agent systems not only adhere to high standards of creative responsibility but also foster trust and confidence among users, collaborators, and audiences.

We acknowledge the potential connection between our musical agent systems and music copyright in case users may be involved in training on existing corpora created by other musicians. However, even in this scenario, our approach closely aligns with the typical practices of DJs who sample, mix, and manipulate musical works, ensuring proper attribution through track identification for each song, as well as with the approaches of electroacoustic musicians working in musique concrète, which heavily incorporate sound design and sampling techniques. As a result, the ethical considerations surrounding our system are far distant and different from recent ongoing cases of alleged copyright infringement and exploitation, such as those involving Sono and Udio in 2024, without acknowledging musicians with the training big data for their AI models. We are committed to opposing any form of copyright infringement or exploitation of musicians. Furthermore, most users of our system are techno-fluent musicians, composers, and sound artists who collaborate with AI systems to expand and explore their artistic and creative potential by training their composition and corpora.

Aside from the above points, training AI models on smaller datasets is more environmentally friendly than training on large datasets because it requires fewer computational resources and consumes less energy, resulting in a smaller carbon footprint. Our musical agent systems do not require external GPUs for training; they have been effectively trained using only the Apple M2 CPU found in a MacBook Pro or even older MacBook models equipped with an Intel core i7 CPU. Small data models lead to quicker, more efficient training without GPUs or cloud computing systems, reducing energy consumption and aligning with efforts to combat climate change. Additionally, this approach encourages the development of optimized, energy-efficient algorithms and minimizes the need for frequent retraining. Focusing on small, high-quality datasets also promotes targeted AI development and addresses ethical concerns related to data privacy. Overall, using small datasets supports sustainable AI development by reducing environmental impact and fostering responsible technological innovation.

In summary, our musical agent systems leverage the ethical and creative benefits of small, personalized datasets, offering a more transparent, sustainable, and customizable approach to generative music AI systems. This method not only safeguards the rights and contributions of individual musicians but also ensures that the generated music is innovative, respectful, and aligned with the ethical standards of contemporary artistic and musical practice.

\section{Detailed Explanation of Musical Agent MASOM}
We provide a comprehensive explanation of the latest version of the MASOM musical agent we extended, drawing on references to the closely related MACAT and MACataRT agents, which share similar architecture and concepts.

MASOM aims to create and generate music or a part of a music excerpt, mainly to assist music composition and improvisation in real time. It is designed to accommodate experimental music composition and free improvisation.

\subsection{Interface of MASOM}
\begin{figure}[htbp]
  \centering
  \includegraphics[width=1.0\columnwidth]{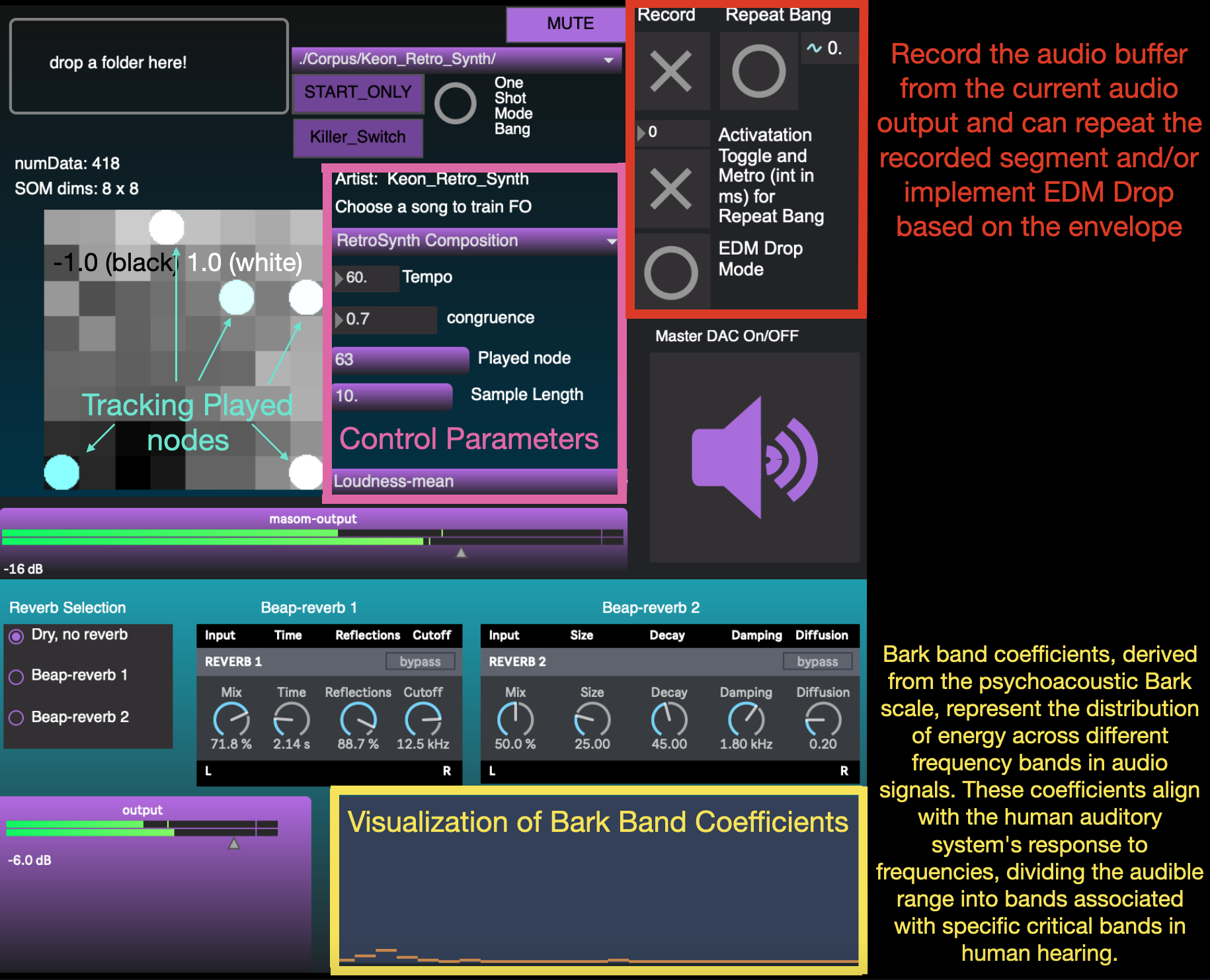}
  \caption{The latest MASOM interface extended by the authors.}
\label{fig:MASOM-Interface}
\end{figure}

Figure \ref{fig:MASOM-Interface} displays the current interface of MASOM extended by the authors. The tempo in the interface is different from the audio corpus's tempo but refers to the playing speed of nodes in the MASOM. The congruence parameter can control the probability (0.0-1.0) of the played node's forward and backward jumps in MASOM. If the congruence is 1.0, MASOM will eventually repeat the same node. If the congruence is 0.0, MASOM tends to play nodes more electrically than neighbouring nodes given the previous node. The bark band coefficients are visualized based on the MASOM's audio output for analysis.

\begin{figure}[htbp]
  \centering
  \includegraphics[width=1.0\columnwidth]{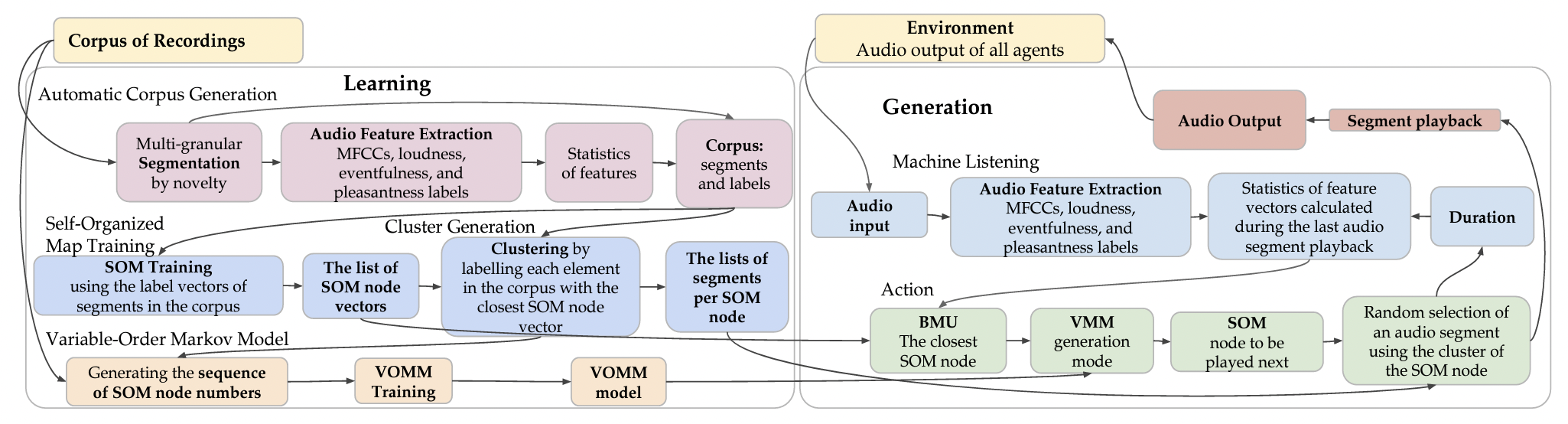}
  \caption{Diagram of original MASOM architecture, as presented in the original MASOM publication.}
\label{fig:MASOM-Architect}
\end{figure}

\subsection{Analysis for MASOM Architecture}

Figure \ref{fig:MASOM-Architect} shows the architecture of MASOM from the original paper. It is important to note from the Figure that the current MASOM architecture is not equipped with the VOMM (Variable-Order Markov Model) but with Factor Oracle (FO). FO in generative music is an algorithmic tool that leverages pattern recognition and a directed acyclic graph structure to identify, store, and utilize recurring musical patterns. It contributes to the generation of music by allowing for the exploration and manipulation of learned patterns to create new and coherent musical sequences.

MASOM is analyzed for acting in an environment based on the conventional AI agent framework:
\begin{enumerate} 
    \item Prior knowledge: The audio recordings (chosen by users) are used to train by applying AI algorithms, and the audio would be input data for this model. So, the musical agent should be pre-trained first (offline training) to use it during live performance. 
    \item Past actions: Automatic music generation in the agent is based on the training data (in audio recordings), and the recordings are converted to multiple audio segments using the multi-granular segmentation based on the Fast Fourier Transform algorithm. The musical agent learns to generate music based on musical memory (from self-organizing map nodes) and statistical learning (learning through the connectivity of each SOM node). The training process is based on past actions, in this case, past audio segments generated by the agent, and it would generate the next node based on past played nodes (non-episodic).
    \item Goal \& values: The goal of the agent is automatic music composition and free improvisation for original music content based on the training data’s style of music. SOM is an artificial neural network that represents, visualizes, and clusters high-dimensional input data in a 2-dimensional topology based on each calculated feature vector. For each SOM node value, SOM clusters input data using the Best Matching Unit (BMU) node, and each square node displays calculated feature vectors. The SOM topology is normalized between -1.0 (black) and 1.0 (white).
    \item Observations: The agent can efficiently listen to a massive amount of music by extracting and analyzing audio features as part of the machine listening module. The agent partially sees the environment because it can perceive the current status (SOM nodes are visualized in real-time) and can control musical parameters to generate different outputs. However, the musical agent cannot control the stochastic nature of the machine learning environment, so the output would be different even if the musical parameters are the same. \\
    \textbf{Internal structure:}
    \begin{enumerate}
        \item Architecture and program: MASOM uses a music programming environment called MAX/MSP/Jitter. The musical agent uses a cognitive architecture based on the multi-agent system. Moreover, the agent is considered a sonic software agent rather than a physical agent, such as a musical robot. 
        \item Knowledge of the environment: MASOM’s environment is based on training data (in audio recordings) and extracted audio features from the training data.
        \item Reflexes: The musical agent has non-reflex actions because it executes an action based on training data (fixed audio corpus by a user), and the agent should be pre-trained before the performance. 
        \item Goals: The goal is automatic music composition and free improvisation for original music content based on the training data’s style of music. However, knowing a goal is challenging, considering MASOM is a musical (creative) agent, so it is difficult to define a problem, unlike most problem-solving agents. Moreover, music tends to be subjective for evaluation, so there is no universal method for evaluating musical metacreation tasks. However, there are use cases in which the agent can improvise during live performances.
        \item Utility functions: The agent's architecture consists of self-organizing maps (musical memory, visualization, unsupervised learning), Variable Markov models and Factor Oracle for learning a musical structure and affective computing for machine listening based on the Circumplex model of affect.
    \end{enumerate}
    \textbf{Details of the environment:}
    \begin{enumerate}
        \item Accessibility \& Determinism: MASOM environment is accessible only for training data based on a specific format of audio recordings (.wav, 16-bit, and 44.1 kHz), not fully accessible for the other data formats, such as MIDI, symbolic sheet music, and non-waveform audio formats. The environment is non-deterministic because the agent cannot recognize the exact state of the world after the agent’s action. Moreover, the generated audio output by the agent would be different in every iteration because of the stochastic nature of the machine learning algorithms.
        
        \item Episodes: In the environment, the training process uses statistical learning based on past actions (the composition of training data). In this case, all audio segments should be analyzed and learned by the agent, and it would generate the next segment based on past segments (non-episodic). The current episode (the current SOM node) is affected by the past episodes (the past SOM nodes). 
        
        \item Dynamic/Static \& Discrete/Continuous: MASOM is static because the architecture has no hierarchy. Audio recordings (training data) are transformed into audio segments (symbolic version of audio recordings, non-continuous audio data), so the agent environment is discrete.
    \end{enumerate}

\end{enumerate}
\subsection{Machine Listening System in MASOM}

The machine listening module incorporates audio feature extraction, affect estimation and statistics calculation. The affect estimation algorithm in the generation module is an online version. MASOM computes audio feature statistics during the sample duration played by the musical action module. Upon triggering a new sample, the machine listening module resets all statistics. It produces a 31-dimensional vector (audio and affective features) for the musical action module.

The musical action module determines distances between the machine listening module's vector and the agent's SOM nodes to identify the Best Matching Unit (BMU), representing the current perceived musical state. The BMU is then sent to FO, where a BMU history is stored to establish context. Using this context, FO predicts the next SOM node to be played. Each SOM node signifies a cluster of audio segments. When the preceding sample playback concludes, the musical agent utilizes the predicted SOM node to select a cluster of audio segments. In the final step, the agent randomly selects a sample within the cluster node to generate the audio output.

The current version of the machine listening system extracts five low-level audio features, and two affective features (Valence and Arousal) are calculated based on the extracted audio features based on the automatic soundscape affect recognition. All audio features are computed with a window size of 1024 samples (duration: 23ms) and a hop size of 512 samples (duration: 12ms).

The extracted audio features are as follows: 

\begin{itemize}
    \item Mel-Frequency Cepstral Coefficients (MFCC): The computation of MFFC involves merging the Mel-frequency scale with a specific frequency spectrum calculation known as cepstrum. The Mel-frequency scale corresponds to the critical bands of human hearing. At the same time, the cepstrum represents the discrete cosine transform (DCT) of the logarithm of the spectrum obtained through the Fast Fourier Transform (FFT). Our calculation yields 13 MFCCs, excluding the zero coefficient. The removal of MFCC0, which represents energy or DC offset, is part of the process.

    \item Loudness: From a psychoacoustical perspective in music, loudness is the subjective perception of a sound's intensity, considering factors like frequency, duration, and sound pressure level. It accounts for the human auditory system's sensitivity to different frequencies, with lower frequencies potentially perceived differently than higher ones, even with equal physical intensity. Psychoacoustic models aim to quantify the nuanced relationship between physical sound characteristics and how humans perceive loudness.

    \item Spectral flatness: Spectral flatness, denoted as the ratio between the geometric mean and the arithmetic mean of the energy spectrum, indicates the noisiness relative to the sinusoidality of the spectrum. To assess this, we calculate the spectral flatness in four frequency bands: 250-500Hz, 500-1000Hz, 1000-2000Hz, and 2000-4000Hz. In Equation 2a, the Spectral-Flatness-1-mean represents the moving average of the computed spectral flatness over the 250-500Hz band.

    \item Perceptual Spectral Decrease: Perceptual spectral decrease in the psychoacoustic context of music refers to the phenomenon where the human auditory system becomes less sensitive to high-frequency spectral components over time. This perceptual shift influences how we perceive and process the frequency content of a musical sound or phrase as it progresses. The concept is crucial in understanding our subjective music experience, impacting factors like timbre perception, sound quality, and tonal balance. In practical terms, it guides audio engineering and music production considerations to align with human auditory perception.

    \item Perceptual Tristimulus: Perceptual tristimulus in psychoacoustical context denotes three perceptual attributes—loudness, pitch, and timbre. This concept aims to capture and represent how the human auditory system perceives and distinguishes various musical sounds, providing a framework for quantifying and analyzing the perceptual dimensions of sound in a manner aligned with human auditory experiences. These parameters contribute to understanding how different musical elements shape the overall perception of a piece of music.

\end{itemize}

The valence and arousal are computed based on the equations (1) and (2) (STD= standard deviation): 
\begin{equation}
\begin{split}
Valence = -0.169 + 
(0.061 * \text{Loudness\_Mean}) + \\
(0.588 * \text{Spectral\_Flatness\_1\_Mean}) + \\
(0.302 * \text{MFCC\_1\_STD}) + \\
(0.361 * \text{MFCC\_5\_STD}) - \\
(0.229 * \text{Perceptual\_Spectral\_Decrease\_STD})
\end{split}
\label{equation-1}
\end{equation}

\begin{equation}
\begin{split}
Arousal = -1.551 + 
(0.060 * \text{Loudness\_Mean}) + \\
(0.087 * \text{Loudness\_STD}) + 
(1.905 * \text{Perceptual\_Tristimulus\_2\_STD}) + \\
(0.698 * \text{Perceptual\_Tristimulus\_3\_Mean}) + 
(0.560 * \text{MFCC\_3\_STD}) - \\
(0.421 * \text{MFCC\_5\_STD}) + 
(1.164 * \text{MFCC\_11\_STD})
\end{split}
\label{equation-2}
\end{equation}

\section{Interface for Musical Agent Systems}
\subsection{Interface of the MACAT System}
\begin{figure}[htbp]
  \centering
  \includegraphics[width=0.7\columnwidth]{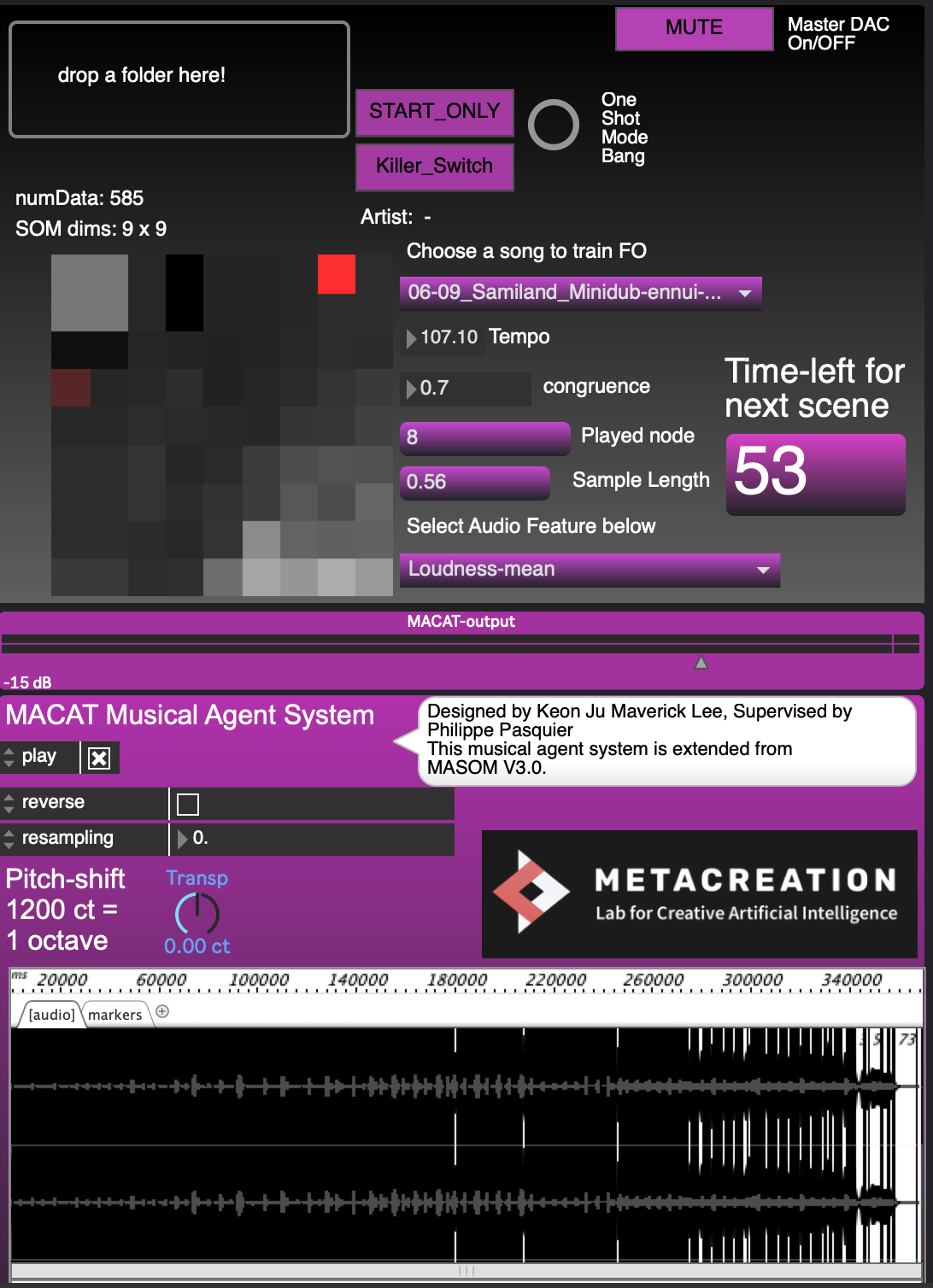}
  \caption{The interface of MACAT system.}
\label{fig:MACAT-Interface}
\end{figure}
Figure \ref{fig:MACAT-Interface} presents the interface of the MACAT system. The upper section of the interface consists of self-organizing maps and its visual representation. In the visual representation, the red block with higher luminance indicates the node currently being played by MACAT, while the red block with lower luminance signifies the previously played node.

The area above the visual representation displays the prompt "Drop a folder here!" which allows the user to load a pre-trained model for the agent. It also presents the values "numData: 99" and "SOM dims: 4 x 4," indicating that the pre-trained model contains 99 audio segments. These segments are organized using a self-organizing map (SOM) with dimensions of 4 by 4. Each node within the SOM encompasses multiple audio segments grouped according to a clustering algorithm used for representation. Right next to the visual representation are parameters such as tempo and congruence that control the agent's behaviour.

Additionally, the system outputs the currently played node number, the corresponding audio sample's duration (in seconds) and the trained corpus's artist and song names in text format. The user has the option to select the specific audio feature to be visually represented, and the agent also informs the time left (in seconds) before 60 seconds for the next music scene, which is helpful information during the performance, consisting of multiple music scenes. In the visual representation, lower feature values (approaching -1.0) are indicated by black colour blocks, while higher values (approaching 1.0) are represented by white colour blocks. Aside from that, the agent has start/stop, restart, mute, master audio system on/off buttons, and one-shot mode bang blinks every time each node is played.

The lower section of the interface integrates a concatenative sound synthesis module and a visualization of the segmented training corpus. Markers indicate the segmentation points, while the currently active audio segment is highlighted in green. Additionally, the interface presents control parameters for the sound synthesis module, including options for playback, reverse, resampling, and pitch shifting, which are integral to our musical practice. The playback function triggers the agent’s sound output, while the reverse function inverts the audio sample during playback. The resampling parameter simultaneously adjusts the playback speed and alters the pitch of the audio sample. The pitch shift parameter operates independently of the playback speed of the audio segment, utilizing units measured in cents (abbreviated as "ct"). In musical terms, a pitch shift of 1200 cents corresponds to an interval of one octave on the musical scale.

\subsection{Interface of the MACataRT System}
\begin{figure}[htbp]
  \centering
  \includegraphics[width=1.0\columnwidth]{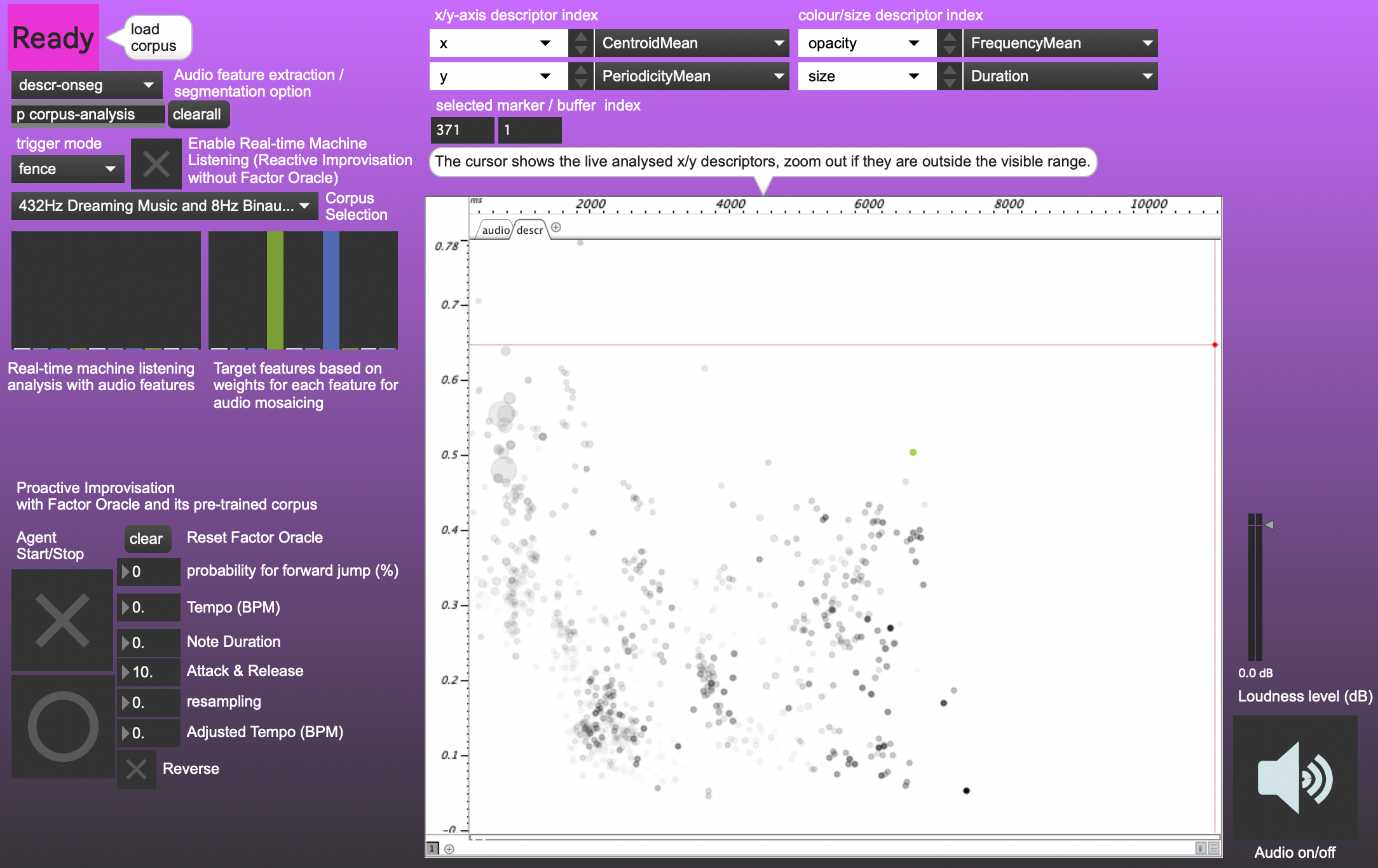}
  \caption{The interface of MACataRT system.}
\label{fig:MACataRT-Interface}
\end{figure}
Figure \ref{fig:MACataRT-Interface} illustrates the interface of the MACataRT system, an extension of IRCAM’s CataRT system. Central to the interface is a two-dimensional scatter plot, which visualizes audio segments based on two chosen audio features: CentroidMean on the x-axis and PeriodicityMean on the y-axis. Additional audio features, FrequencyMean and Duration, are represented through the opacity and size of the plotted points, respectively. Higher FrequencyMean values are indicated by increased opacity, while larger point sizes correspond to longer durations of the audio segments. Alternative audio features can be easily selected, and colour coding may replace opacity for visual differentiation. The interface also tracks the currently selected marker index and buffer index, with the selected audio segment highlighted as a green dot on the scatter plot.

The upper-left section of the interface displays the "Ready" status instead of "Load" once the corpus has been loaded for analysis and segmentation into multiple samples, visualized within the scatter plot. For audio feature extraction and segmentation, the system offers several options via the Max/MSP MuBu library, each providing a range of adjustable parameters tailored to the user's specific requirements in their musical practice. In addition, users can access the trigger mode selection, which offers various options within CataRT, including "bow", "fence", "beat", "loop", "beatmove", "loopmove", and "cont". These modes provide flexibility to accommodate a range of musical scenarios, allowing users to tailor the system’s behaviour to their specific creative requirements.

The lower-left section of the interface provides control parameters for the system's temporal model, which is based on the factor oracle. Users can initiate, stop, and reset the musical agent system, as well as adjust the probability of forward jumps. For instance, a probability setting of 100\% ensures the factor oracle will move exclusively to forward nodes, while a 50\% setting results in an equal likelihood of forward and backward jumps. A 0\% probability restricts the system to only backward jumps. As the factor oracle's suffix tree exhibits stochastic behaviour, outcomes are inherently variable and not fully predictable. 

The tempo parameter determines the rate at which the agent progresses to the next node, with note durations corresponding to the length of each audio segment. The attack and release parameters facilitate smooth transitions by applying fade-in and fade-out effects, preventing audio clipping and abrupt shifts. Similar to the MACAT system, the reverse parameter allows each audio segment to be played in reverse while the resampling parameter adjusts both pitch and playback speed. Our musical practice with the system has shown that calculating the adjusted tempo when using the resampling parameter is essential to maintaining a coherent rhythm, as a properly aligned tempo prevents disruptions in rhythm when altering the resampling settings.

The MACataRT system integrates a temporal model utilizing the factor oracle, a pattern-matching suffix automaton. This enables the agent to learn and adapt to the musical and timbral characteristics of the training corpus. The system also allows the agent to analyze and interact with musical input by incorporating a machine listening module and an audio mosaicing technique.

In music improvisation, the audio mosaicing technique can be employed to dynamically reassemble fragments of pre-recorded audio in response to real-time musical input. For example, during an improvisational performance, a system utilizing audio mosaicing might analyze live input from a performer, such as a melodic or rhythmic pattern, and respond by selecting and recombining audio segments from a pre-existing sound corpus. These fragments are chosen based on acoustic features—such as pitch, timbre, or rhythm—that match or complement the live input. This process enables the system to generate novel, contextually responsive sonic textures, allowing the performer to interact with an evolving soundscape that incorporates both the pre-recorded material and the live improvisation. Through this interaction, the system actively participates in the improvisational process, facilitating new forms of creative expression and dialogue between human performers and the machine.

The MACataRT system supports two modes of operation: reactive improvisation and proactive improvisation. The system does not utilize the factor oracle temporal model in reactive improvisation. Instead, the agent generates audio in response to the human musician's input, using audio mosaicing techniques based on selected audio features determined by their corresponding feature weights. The feature weights corresponding to specific audio characteristics are visualized through histograms, as depicted in Figure \ref{fig:MACataRT-Interface}. Adjacent to this, an additional histogram provides a real-time representation of the machine listening analysis based on the input audio. Typically, users select two target features from the upper part of the interface, which are displayed along the x and y axes, with CentroidMean and PeriodicityMean shown as the selected features, respectively.

For proactive improvisation, the agent employs the temporal model known as the factor oracle, which necessitates a prior training phase. This training involves an offline machine listening process to construct the factor oracle by identifying patterns within the sequence of audio segment indices. These indices are derived from the audio mosaicing process used during reactive improvisation. Once trained, the factor oracle enables the agent to improvise by generating sequences based on the learned patterns while dynamically adjusting its performance in real-time through the manipulation of control parameters.

\end{document}